\documentclass[aps,prb,a4paper,twocolumn,floatfix,showkeys,amsmath,amssymb,nofootinbib,nobibnotes,altaffilletter]{revtex4}
\usepackage[pdftex]{hyperref}

\usepackage[T1]{fontenc}
\usepackage[latin1,utf8]{inputenc}
\usepackage[english]{babel}

\usepackage{setspace}

\usepackage{amsmath}
\usepackage{bm}

\usepackage{graphicx}
\usepackage{pslatex}
\usepackage{xspace}
\usepackage{subfigure}
\usepackage{float}

\usepackage{booktabs}

\usepackage{ragged2e}
\usepackage[labelfont=bf, format=plain, indention=0.5cm, textfont=small, skip=4pt, justification=RaggedRight, tableposition=top, figureposition=bottom ]{caption}

\newcommand{\degree}{$^{\circ}$}

\renewcommand{\eqref}[1]{Eq.~(\ref{eq:#1})}

\floatstyle{boxed}
\newfloat{copyrightfloat}{thp}{lop}

\begin{document}

\begin{copyrightfloat}
\raggedright
The peer reviewed version of the following article has been published in final form at  J. Am. Chem. Soc., 2011, 133, 9088, doi: \href{http://dx.doi.org/10.1021/ja2025432}{10.1021/ja2025432}.
\end{copyrightfloat}

\preprint{}

\title{Triplet Exciton Generation in Bulk-Heterojunction Solar Cells based on Endohedral Fullerenes}

\author{Moritz \surname{Liedtke}}\email{Moritz.Liedtke@Physik.Uni-Wuerzburg.de}
\affiliation{Bavarian Center for Applied Energy Research (ZAE Bayern), 97074 W{\"u}rzburg, Germany}

\author{Andreas \surname{Sperlich}}
\author{Hannes \surname{Kraus}}
\author{Andreas \surname{Baumann}}
\author{Carsten \surname{Deibel}}
\affiliation{Experimental Physics VI, Julius-Maximilian University of W{\"u}rzburg, 97074 W{\"u}rzburg, Germany}

\author{Maarten J. M. \surname{Wirix}}
\affiliation{Laboratory for Materials and Interface Chemistry, Eindhoven University of Technology, 5600MB Eindhoven, Netherlands}

\author{Joachim \surname{Loos}}
\affiliation{Laboratory for Materials and Interface Chemistry, Eindhoven University of Technology, 5600MB Eindhoven, Netherlands}
\altaffiliation{School of Physics and Astronomy, Kelvin Nanocharacterisation Centre, University of Glasgow, G12 8QQ Glasgow, Scotland}

\author{Claudia M. \surname{Cardona}}
\affiliation{Luna Innovations Incorporated, 521 Bridge Street, Danville, VA 24541, USA}

\author{Vladimir \surname{Dyakonov}}\email{Dyakonov@Physik.Uni-Wuerzburg.de}
\affiliation{Experimental Physics VI, Julius-Maximilian University of W{\"u}rzburg, 97074 W{\"u}rzburg, Germany}
\affiliation{Bavarian Centre for Applied Energy Research (ZAE Bayern), 97074 W{\"u}rzburg, Germany}

\date{March 21, 2011}


\begin{abstract}
Organic bulk-heterojunctions (BHJ) and solar cells containing the trimetallic nitride endohedral fullerene 1-[3-(2-ethyl)hexoxy carbonyl]propyl-1-phenyl-Lu$_{3}$N@C$_{80}$ show an open circuit voltage ($V_\text{OC}$) 0.3~V higher than similar devices with [6,6]-phenyl-C[61]-butyric acid methyl ester (PC$_{61}$BM). To fully exploit the potential of this acceptor molecule with respect to the power conversion efficiency (PCE) of solar cells, the short circuit current ($J_\text{SC}$) should be improved to become competitive with the state of the art solar cells. Here we address factors influencing the $J_\text{SC}$  in blends containing the high voltage absorber Lu$_{3}$N@C$_{80}$-PCBEH both in view of photogeneration but also transport and extraction of charge carriers. We apply optical, charge carrier extraction, morphology and spin sensitive techniques. In blends containing Lu$_{3}$N@C$_{80}$-PCBEH we found two times weaker photoluminescence quenching, remainders of interchain excitons and mostly remarkable, triplet excitons formed on the polymer chain, which were absent in the reference P3HT:PC$_{61}$BM blends. We show that electron back transfer to the triplet state along with the lower exciton dissociation yield due to intramolecular charge transfer in Lu$_{3}$N@C$_{80}$-PCBEH are responsible for the reduced photocurrent.
\end{abstract}


\maketitle

\section{Introduction}


Organic solar cells offer a chance towards low-cost and clean energy supply in the near future. Best laboratory scale devices to date reached confirmed power conversion efficiencies (PCE) of 8.3~\%~\cite{Green:2011hq}. To make these cells competitive in the solar market, further improvements of the PCE is needed~\cite{Deibel:2010do}. Development of new materials is a prequisit for successful development of the organic photovoltaic (OPV) technology. Most of the activities to date are focused on low-band gap materials and donor-acceptor copolymers~\cite{Liang:2010jb,Peet:2007fx}. Being efficiently generated, singlet excitons must be splitted into separate electrons and holes. Fullerenes are used as acceptors in BHJ solar cells due to their excellent electron acceptor properties auxiliary for singlet exciton dissociation at the donor--acceptor interfaces and their ability to form electron transport pathways across the semiconductor layer. Exciton splitting is, however, accompanied by the loss of potential energy of electrons resulting in relatively low $V_\text{OC}$, compared to the exciton generation energy of about 2~eV. $V_\text{OC}$'s in the range of 0.55-0.65~V are typical for PC$_{61}$BM, PC$_{71}$BM or diphenyl-methanofullerenes DPM-12-C$_{60}$~\cite{Dennler:2009gg,Ameri:2010ko,Riedel:2004dy,Riedel:2005ec,SanchezDiaz:2010iu} based organic solar cells using P3HT as donor material.

C$_{80}$ fullerene derivatives used as acceptors attracted a lot attention recently due to their potentially high $V_\text{OC}$~\cite{Ross:2009eo,Ross:2009kb,Kooistra:2006gp}. The enhanced open circuit voltage we found in our samples comes along with a reduced $J_\text{SC}$ resulting in a PCE comparable to that of common P3HT:PC$_{61}$BM cells made by us (see below) and others~\cite{Yang:2005bt,Dennler:2009gg}. This may be either understood in terms of electron back transfer (EBT) to the polymer triplet state due to the high LUMO level of Lu$_{3}$N@C$_{80}$-PCBEH~\cite{Veldman:2009cw}, or due to electron transfer from the endohedral metal cluster to the cage, thereby changing the electronic properties of the molecule and thus its electron affinity~\cite{Kobayashi:1995co,Xu:2011dw,Iezzi:2002et,Iiduka:2005hw}.

We found a photoluminescence quenching lower than for P3HT:PC$_{61}$BM, and mostly intriguing triplet states in the blend of P3HT:Lu$_{3}$N@C$_{80}$-PCBEH. The later state seems to be formed either due to EBT or by direct inter system crossing (ISC) from the polymer singlet state by  not dissociated excitons.

\section{Results and discussion}

\begin{figure}[ht]
 \centering
	\includegraphics{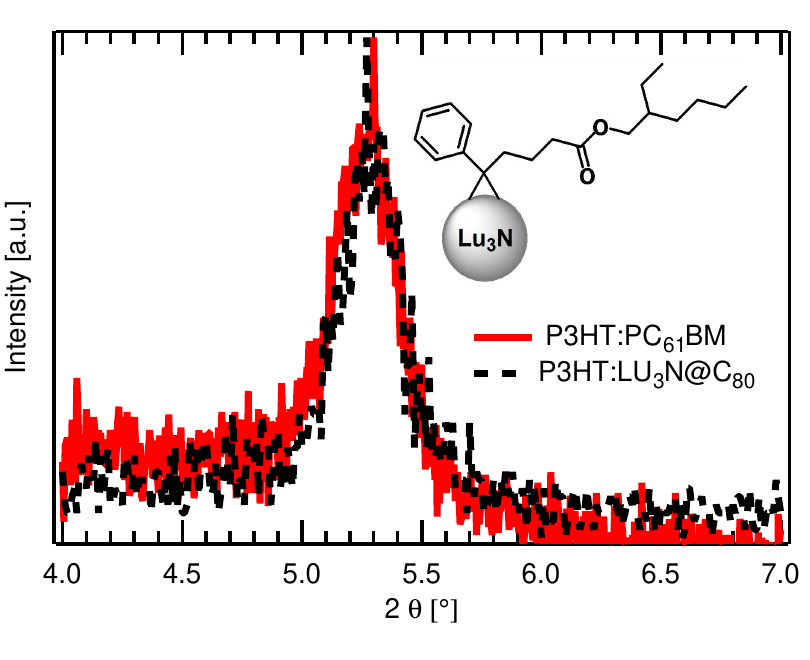}
	\caption{XRD spectra of P3HT:Lu$_3$N@C$_{80}$-PCBEH and P3HT:PC$_{61}$BM organic solar cells. Both samples show a peak at around 5.3\degree. This corresponds to a P3HT (100) lattice distance of 16.6~\AA. Inset: Structure of Lu$_{3}$N@C$_{80}$-PCBEH.}
	\label{fig:XRD}
\end{figure}

\begin{figure}[ht]
	\centering
	\includegraphics[height=6.4cm]{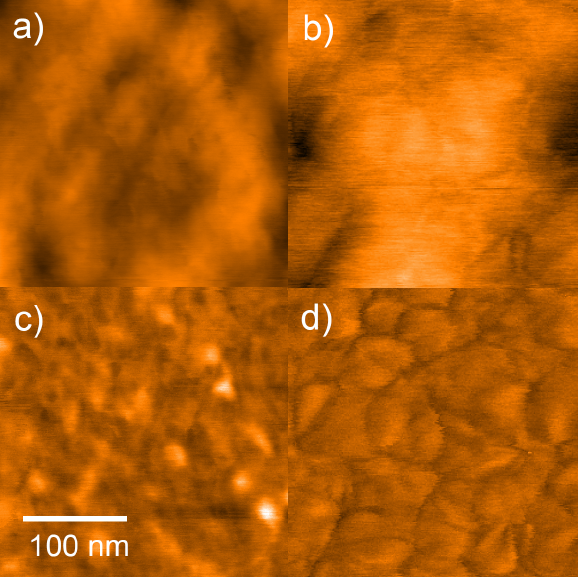}
	\caption{AFM images of the donor:acceptor blends of the organic solar cells. a) and b) show the height image of P3HT:Lu$_3$N@C$_{80}$-PCBEH and P3HT:PC$_{61}$BM organic solar cells, c) and d) the corresponding phase images. The areas are 250x250~nm, surface roughness of both samples is below 0.7~nm.}
	\label{fig:AFM}
\end{figure}

\begin{figure}[ht]
	\centering
	\includegraphics[height=6.4cm]{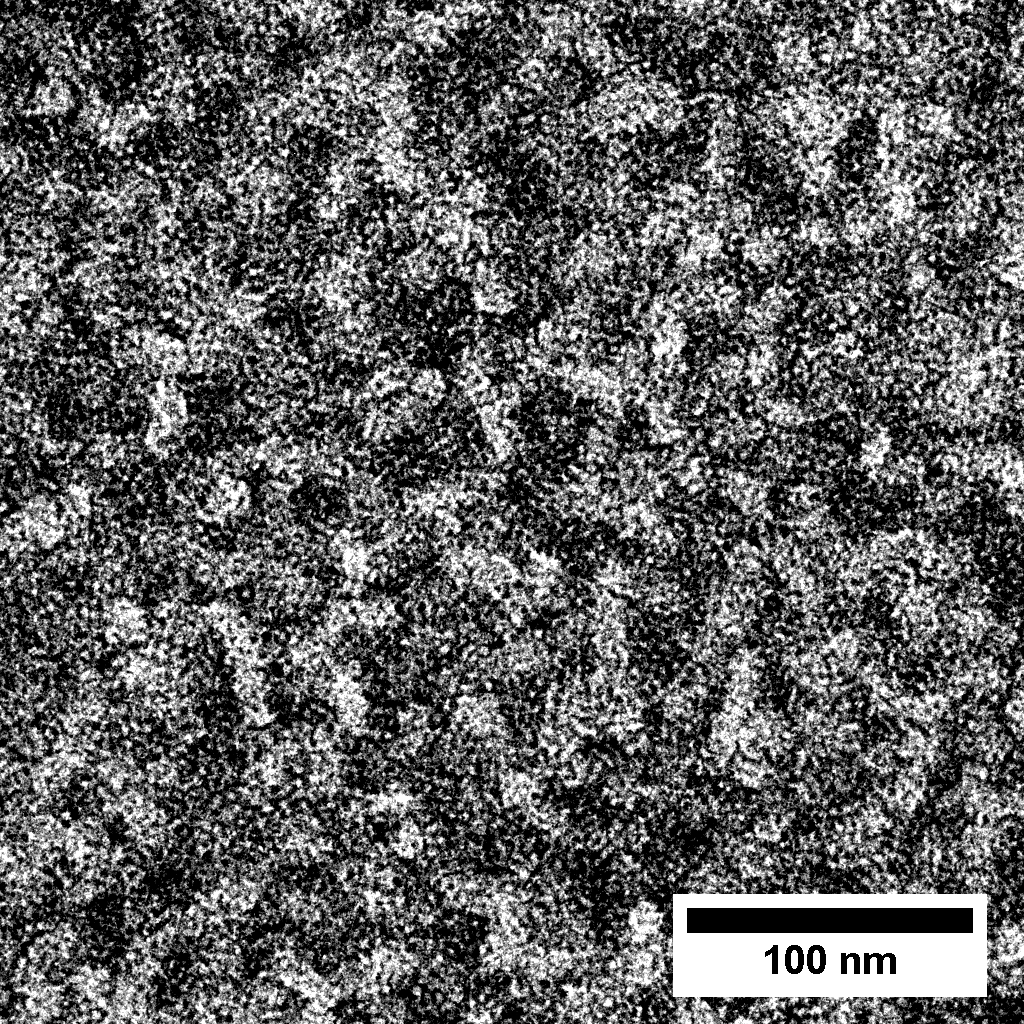}
	\caption{TEM image of the donor-acceptor phase for annealed P3HT:Lu$_3$N@C$_{80}$-PCBEH blend. Intermixing in the scale of 15~nm was found. This is comparable to pristine P3HT:PC$_{61}$BM blends~\cite{Yang:2005bt}.}
	\label{fig:TEM}
\end{figure}

The exciton dissociation efficiency is related to the relative electron affinity (LUMO levels) of donor and acceptor as well as to the issue of phase separation. To probe the morphology of the solar cell's active layer we applied X-ray diffraction (XRD), atomic force microscopy (AFM and transmission electron microscopy (TEM). XRD showed peaks at 5.3\degree~(Fig.~\ref{fig:XRD}), corresponding to an interplanar distance of 16.6~\AA~of the P3HT chains in a lattice for both blends. This is in good agreement with literature data~\cite{Prosa:1992vy,Chen:1995gc,Vanlaeke:2006gv} regarding the lattice distance of P3HT stacks. The graph also includes the structure of the molecules side chain and a cyclic voltammogram of Lu3N@C80-PCBEH in solution. Details about the internal fullerene structure can be found elsewhere~\cite{Stevenson:2002hp}. AFM~(Fig.~\ref{fig:AFM}) revealed smooth surfaces with a roughness below 0.7~nm for both blends. The phase images show a more diffuse pattern for the P3HT:Lu$_3$N@C$_{80}$-PCBEH blend compared to the reference cell, possibly indicating intermixing on a finer scale for P3HT:Lu$_3$N@C$_{80}$-PCBEH. TEM measurements showed an intermixing with domain sizes of about 15~nm and long polymer fibers~(Fig.~\ref{fig:TEM}). This is comparable with the intermixing in a pristine P3HT:PC$_{61}$BM blend as seen in previous measurements using the same technique~\cite{Yang:2005bt}.

\begin{figure}[ht]
	\centering
	\includegraphics{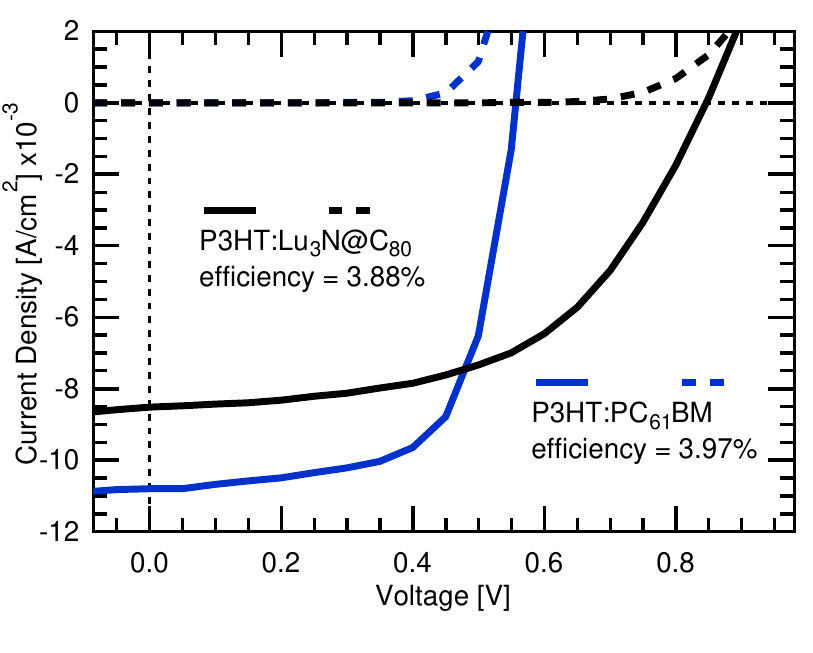}
	\caption{J--V characteristics of a P3HT:Lu$_3$N@C$_{80}$-PCBEH 1:1 bulk-heterojunction solar cell ($V_\text{OC}$~=~835~mV, $J_\text{SC}$~=~8.52~mA/cm$^2$, fill factor of 54~\%) and a 1:0.8 P3HT:PC$_{61}$BM cell ($V_\text{OC}$~=~560~mV, $J_\text{SC}$~=~10.80~mA/cm$^2$, fill factor of 66~\%) as reference.}
	\label{fig:IVcurve}
\end{figure}

Current--voltage (J--V) characteristics and PCE of BHJ solar cells made of the P3HT:Lu$_3$N@C$_{80}$-PCBEH blend with different ratios (1:4 to 4:1) were measured. By varying the P3HT:Lu$_3$N@C$_{80}$-PCBEH weight ratio, we have found a PCE maximum at a ratio of 1:1 P3HT:Lu$_3$N@C$_{80}$-PCBEH. This weight ratio corresponds to the one of a 1:0.8 P3HT:PC$_{61}$BM solar cell when considering the different densities of the two acceptors. Therefore both blends have a comparable donor:acceptor volume ratio~\cite{Erwin:2002fd,Geens:2004kv}. It is well known that the donor--acceptor ratio plays an important role for exciton, polaron pair dissociation and transport of charges to the contacts, as the domain sizes and therefore distances to interfaces and transport routes change~\cite{Vanlaeke:2006gv,Chirvase:2004ex}.

The J--V curve of a 1:1 P3HT:Lu$_3$N@C$_{80}$-PCBEH solar cell~(Fig.~\ref{fig:IVcurve}) showed a high $V_\text{OC}$ of 835~mV and a short circuit current ($J_\text{SC}$) of 8.52~mA/cm$^2$. Combined with a fill factor of 54~\% this results in a PCE of 3.88~\%. For comparison, an organic solar cell made from 1:0.8 P3HT:PC$_{61}$BM blend had a PCE of 3.97~\% ($V_\text{OC}$~=~560~mV, $J_\text{SC}$~=~10.80~mA/cm$^2$, fill factor 66~\%). With a similar type of endohedral fullerenes a PCE of up to 4.2~\% was recently demonstrated using P3HT as donor~\cite{Ross:2009eo}.\

\begin{figure}[ht]
	\centering
	\includegraphics{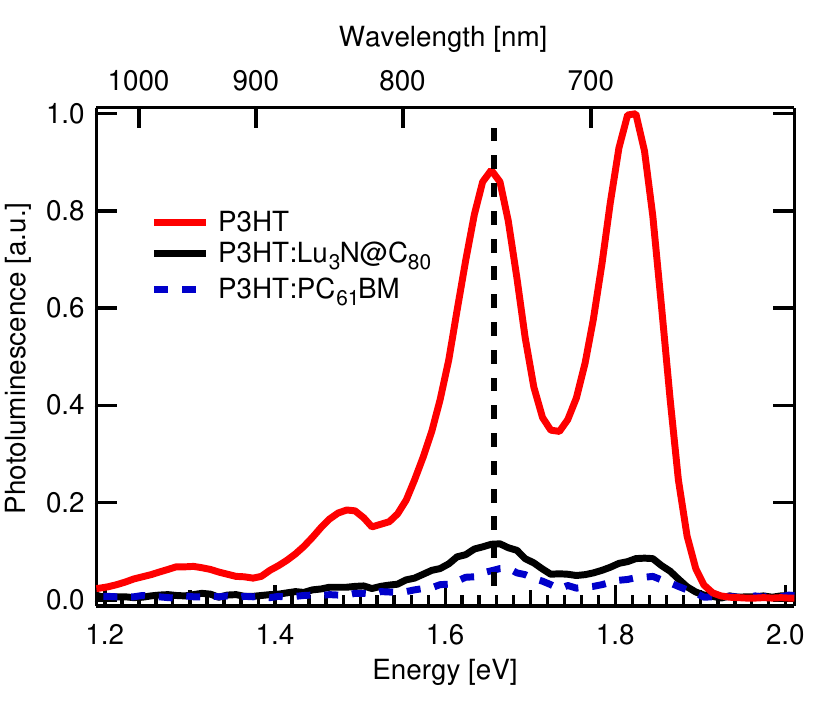}
	\caption{PL of pure P3HT and blends of P3HT:Lu$_3$N@C$_{80}$-PCBEH and P3HT:PC$_{61}$BM. The peak heights at the vertical dashed line are 100~\%, 13~\% and 7.4~\% respectively, all spectra were measured under the same conditions. Indicating a reduced singlet exciton dissociation yield using Lu$_3$N@C$_{80}$-PCBEH as acceptor.}
	\label{fig:PL}
\end{figure}

The first step to generate current in an organic solar cell is the generation of excitons by incoming light, followed by exciton splitting into Coulomb bound charge transfer (CT) excitons or polaron pairs~\cite{Brabec:2008tb}. The quenching of photoluminescence (PL) of the donor when blended with an acceptor is an indicator for the efficiency of the singlet exciton dissociation which may be related, in a next step, to the yield of polaron pair generation. The efficiency of exciton generation in pure P3HT and dissociation in P3HT:acceptor blends after light excitation is well documented~\cite{Dennler:2007kk,Deibel:2010da}. We focus now on the optical properties of the blends. Comparing the PL quenching in blends of P3HT:Lu$_3$N@C$_{80}$-PCBEH and P3HT:PC$_{61}$BM it can be seen that the blending of P3HT with Lu$_3$N@C$_{80}$-PCBEH results in stronger residual PL than with PC$_{61}$BM as acceptor~(Fig.~\ref{fig:PL}).

\begin{figure}[ht]
	\centering
	\includegraphics{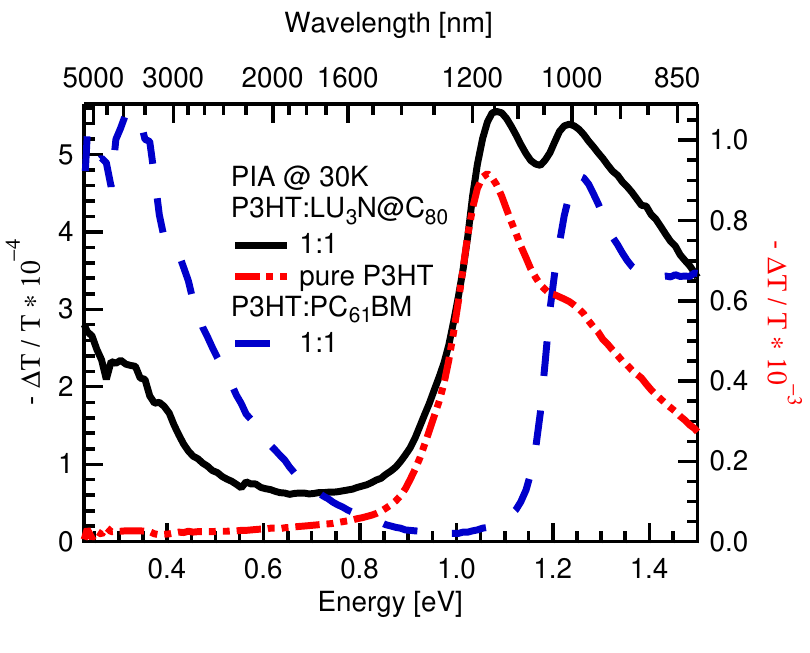}
	\caption{PIA for 1:1 blends of P3HT:Lu$_3$N@C$_{80}$-PCBEH (black line) and P3HT:PC$_{61}$BM (blue dashed line). The graph shows polaronic states at 1.2~eV and 0.3~eV. Additionally, a peak at 1.05~eV indicates interchain-excitons in the P3HT:Lu$_3$N@C$_{80}$-PCBEH blend also seen in pure P3HT (red dash-dotted line).}
	\label{fig:PIA}
\end{figure}

To discriminate between charge and energy transfer, the photoinduced absorption (PIA) technique was applied. It sensitively probes the formation of sub-bandgap states in the near infrared under photoexcitation. The photoinduced absorption in pure P3HT is marked by a broad dominant peak at 1.05~eV showing a shoulder at 1.25~eV~(see Fig.~\ref{fig:PIA}). This spectrum is well established~\cite{Osterbacka:2000vg,Korovyanko:2001gj}. After photo-induced charge transfer between PC$_{61}$BM and P3HT a low energy peak with the maximum at approximately 0.3~eV and a high energy peak at 1.25~eV appear. Both attributed to positively charged polarons on a polymer chain. Note, the PIA peak from radical anion (negative polaron) of the fullerene is overlapping with the radical cation peak at 1.25~eV~\cite{Ramos:2001hh}. Comparing the PIA spectra of P3HT:Lu$_3$N@C$_{80}$-PCBEH with the P3HT:PC$_{61}$BM blend, we find an additional peak at 1.05~eV absent in blends of P3HT:PC$_{61}$BM. This peak was found by Korovyanko et al. and explained as to neutral excitations, such as CT excitons or polaron pairs on the P3HT~\cite{Korovyanko:2001gj}. The peaks at 1.2~eV and 0.3~eV from positive polarons are indicative for a charge transfer that took place in blends of P3HT:Lu$_3$N@C$_{80}$-PCBEH, whereas the peak at 1.05~eV supports the scenario of Coulomb bound polaron pairs, remnant or recovered, in the C$_{80}$ based blends.

\begin{figure}[ht]
	\centering
	\includegraphics{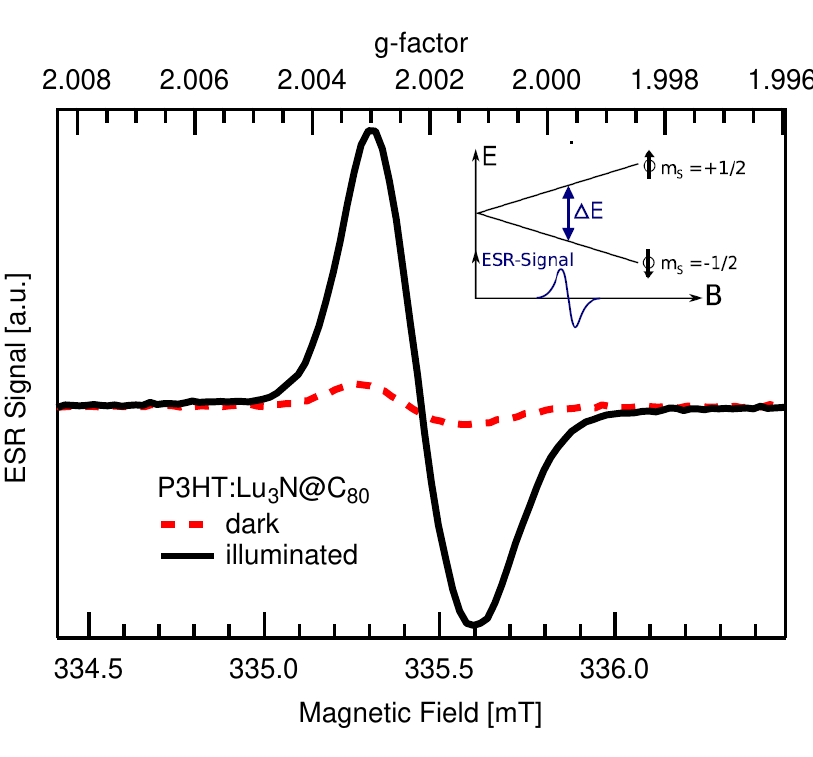}
	\caption{ESR in P3HT:Lu$_3$N@C$_{80}$-PCBEH. Strong radical cation signal at g~=~2.002 is visible when illuminated (solid line), indicating a positive charge on the P3HT chain as a result of a photoinduced charge transfer. A weak signal in the dark is also shown.}
	\label{fig:ESR}
\end{figure}

Further information about the excited species generated in blends is gained from electron spin resonance (ESR) measurements. Pure Lu$_3$N@C$_{80}$-PCBEH revealed a very broad, weak, light-independent ESR signal centered around g~=~2. A positive polaron signal in the P3HT:Lu$_3$N@C$_{80}$-PCBEH blend under illumination can be clearly seen in Fig.~\ref{fig:ESR}. The signal has a g-factor of 2.002 and can be attributed to positive polarons on the P3HT~\cite{Ceuster:2001hc}.

\begin{figure}[ht]
	\centering
	\includegraphics{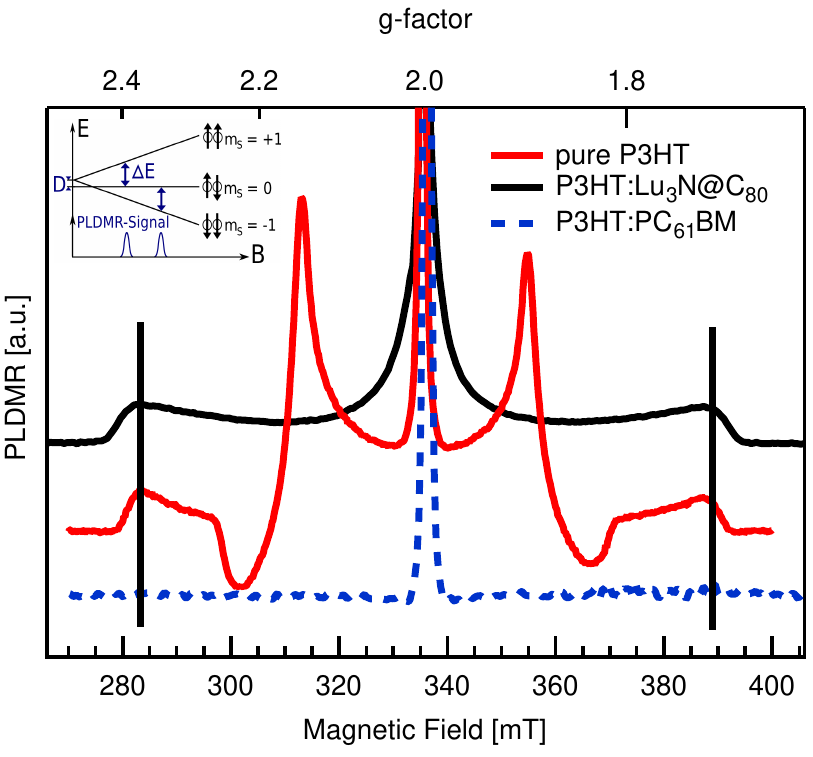}
	\caption{The PLDMR of pure P3HT and OPV blends exhibits a strong narrow signal at g~=~2 corresponding to CT states (polaron pairs). Furthermore, in P3HT a 105~mT wide signature linked with a more localized triplet exciton can be observed. This triplet is quenched by CT in blends with PCBM, while it abides in P3HT:Lu$_3$N@C$_{80}$-PCBEH. The inset shows the microwave induced PLDMR signal generation principle for a D~$\neq$~0 triplet exciton.}
	\label{fig:PLDMR}
\end{figure}

The PLDMR (photoluminescence detected magnetic resonance) method is one of the few facilities to directly prove the existence of triplet excitons, without taking the often indistinct detour over excitation lifetimes. In PLDMR, the microwave induced transitions between Zeeman sublevels of spin-carrying species can be monitored via PL intensity. The so-called full field triplet signature, distinguishable by its wing-like appearance (as can be seen in Fig.~\ref{fig:PLDMR}) is clear evidence for a localized triplet exciton partaking in the excitation layout of a material system. This characteristic signal shape stems from the dipole-dipole interaction \textbf{D} in the spin hamiltonian of the triplet system

\begin{align}
   \mathcal{H} = \textbf{g} \cdot \mu_B B \cdot \hat{S} + \underbrace{\hat{S}^T \textbf{D} \hat{S}}_{Zero field splitting}
\end{align}

leading to the zero field splitting D, as shown in Fig.~\ref{fig:PLDMR} inset.
The figure itself shows the relative change of PL in the blends as well as in the pure P3HT. Triplet exciton signals were only found in pure P3HT and P3HT:Lu$_3$N@C$_{80}$-PCBEH, but not in P3HT:PC$_{61}$BM~\cite{Swanson:1990fu}. As the sensitivity of the method is very high, the intensity of the residual PL in the blends is sufficient for a good signal-to-noise ratio.

\begin{figure}[ht]
	\centering
	\includegraphics{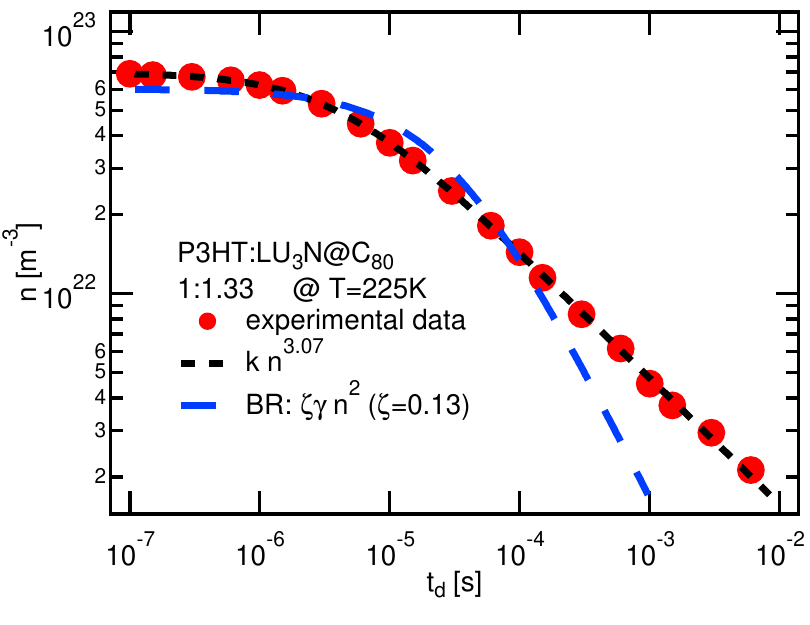}
	\caption{CELIV measurement shows the concentration of extracted charge carriers as function of delay time (red circles) on P3HT:Lu$_3$N@C$_{80}$-PCBEH device. Reduced bimolecular Langevin recombination (blue dashed line) and the recombination with decay order of 3 (black dashed line) are shown. T~=~225~K.}
	\label{fig:CELIV}
\end{figure}

To gather information on the recombination behavior of already separated charge carriers photo-CELIV (charge extraction by linear increasing voltage) experiments were performed. By variation of the delay time between the photo-generation by laser and the extraction of the generated carriers by the voltage ramp, the time dependence of the charge carrier concentration can be studied. With increasing delay time less charge carriers can be extracted due to recombination processes within the active layer. As can be seen in Fig.~\ref{fig:CELIV}), the charge carrier concentration at T~=~225~K can be described best with a bimolecular recombination with carrier concentration dependent (bimolecular) recombination coefficient (black dashed line). This behavior was found previously also in the material system P3HT:PC$_{61}$BM~\cite{Foertig:2009fw,Deibel:2008jj,Juska:2006cc,Shuttle:2008jy}.

The objective of the presented work was to find the explanation for the reduced short circuit current in Lu$_3$N@C$_{80}$-PCBEH based organic solar cells. This will be discussed in terms of exciton generation, charge carrier dissociation and extraction of charge carriers to outer circuits through contacts.

Light harvesting in both blends is assumed to be identically efficient, as light absorption and exciton generation mainly occurs in the polymer phase. To dissociate, singlet excitons have to reach a donor:acceptor interface first. As shown by AFM~(Fig.~\ref{fig:AFM}) and TEM~(Fig.~\ref{fig:TEM}) measurements the intermixing in an annealed blend of P3HT:Lu$_3$N@C$_{80}$-PCBEH is even finer compared to that of an annealed P3HT:PC$_{61}$BM. Therefore the yield of singlet excitons able to reach an interface should be higher for Lu$_3$N@C$_{80}$-PCBEH.\\

The higher residual PL found in blends with Lu$_3$N@C$_{80}$-PCBEH~(Fig.~\ref{fig:PL}), is however a strong indicator that the portion of excitons dissociated is lower compared to PC$_{61}$BM based blends. Morphology measurements ruled out insufficient intermixing of the two components as origin of the stronger residual PL. Thus we assume the intrinsic properties of the Lu$_3$N@C$_{80}$-PCBEH molecule to be responsible for the reduced exciton dissociation yield. They are the high lowest unoccupied molecular orbital (LUMO) level of Lu$_3$N@C$_{80}$-PCBEH (-3.69~eV measured by cyclic-voltammetry) and the high probability of electron transfer from the encased lutetium atoms to C$_{80}$, thereby changing its electron affinity compared to a neutral C$_{80}$ fullerene~\cite{Kobayashi:1995co,Xu:2011dw,Iezzi:2002et,Iiduka:2005hw}.

While the high LUMO level of Lu$_3$N@C$_{80}$-PCBEH on the one hand results in a high $V_\text{OC}$~\cite{Ross:2009eo,Ross:2009kb,Gadisa:2004hw,Brabec:2001fd} it might lead to a higher triplet generation yield on the other hand. Triplet excitons in conjugated polymers can be generated either via intersystem--crossing (ISC) or via CT states (polaron pairs) undergoing electron back transfer (EBT)~\cite{Veldman:2008fx,Dyakonov:1997vx}. ISC occurs when excitons generated on P3HT chains do not undergo a dissociation into polaron pairs on an ultrafast timescale, but instead are converted to a triplet state. EBT results from already separated singlet excitons which do not further dissociate into free electrons and holes. In polymer:fullerene blends, the formation of triplets and forward electron transfer (ET) are competing processes. The ET from polymer to fullerene is considered to be much faster, in the range of few tens of fs, whereas the ISC is in the range of 100~ps~\cite{Cunningham:2008dv}. Once generated, CT states may undergo an EBT, forming triplet excitons on the polymer chain, depending on the relative energetic position of the CT state and the lowest triplet state of the polymer.

In pure P3HT most excited states recombine via photoluminescence while a small number undergoes ISC forming triplet states on the polymer chain, as evidenced by our PLDMR measurements~(Fig.~\ref{fig:PLDMR}). Note we are able to distinguish between triplets formed on the polymer and the fullerene. By adding PC$_{61}$BM to the P3HT the excitons mostly undergo charge transfer and the electrons are transfered to the PC$_{61}$BM with a LUMO level of -4.3~eV~\cite{Scharber:2006gt}. As this level is energetically lower than that of the P3HT triplet state no electron back transfer (EBT) takes place.

Instead triplet excitons can be generated if the CT state lies energetically higher than the lowest triplet state of the polymer. As the LUMO energy of the acceptor influences the energy of the CT state~\cite{Deibel:2010cz}, such situation can be conceivable in blends based on Lu$_3$N@C$_{80}$-PCBEH. This would open a path to form triplet states via EBT and thus reduce free charge carrier generation. Thus the lowest triplet level of the donor sets the upper limit for the CT state energy to prevent EBT. The advantage of a high $V_\text{OC}$ can be contradicted by the loss of charge carriers due to EBT. However, EBT does not explain the higher residual PL in P3HT:Lu$_3$N@C$_{80}$-PCBEH blends compared to P3HT:PC$_{61}$BM.

Electron transfer from the lutetium atoms to the C$_{80}$ cage can account for the reduced singlet exciton dissociation yield. As shown for lanthanum atoms inside a C$_{80}$ fullerene and for lutetium inside a C$_{88}$ fullerene some of the valence electrons of these atoms are transfered to the fullerene~\cite{Kobayashi:1995co,Xu:2011dw}. As lutetium is of the same series as lanthanum, but even heavier and weaker bound valence electrons, intramolecular electron transfer to the C$_{80}$ might be feasible in the case of Lu$_3$N@C$_{80}$-PCBEH. This reduces the electron affinity of Lu$_3$N@C$_{80}$-PCBEH compared to a neutral C$_{80}$. Therefore less singlet excitons are expected to be dissociated at the donor:acceptor interface. These singlet excitons would rather recombine radiatively or undergo ISC.

This scenario is consistent with our photoluminescence and PIA measurements (Fig.~\ref{fig:PL} and Fig.~\ref{fig:PIA}) showing a reduced quenching in the blends with Lu$_3$N@C$_{80}$-PCBEH and also fits the results of the PLDMR measurements~(Fig.~\ref{fig:PLDMR}). Less efficient exciton dissociation yield due to intramolecular charge transfer may explain the reduced photocurrent.

So far we discussed the reduced $J_\text{SC}$ in terms of charge carrier generation yield. On the other hand it may also be due to transport properties, e.g. high recombination. To address this issue we performed photo--CELIV measurements. The results showed a reduced Langevin recombination rate similar to the one typically observed in BHJ devices~\cite{Foertig:2009fw,Juska:2006cc}. The polaron recombination dynamics is  characterized by charge decay order being two in case of Langevin. As shown in Fig.~\ref{fig:CELIV} the recombination order is higher. We assign the tail at longer times to a delayed release of charge carriers not subjected to recombination~\cite{Baumann:2011eo}. Therefore, the reduced photocurrent can be attributed to a reduced photogeneration yield.

\section{Summary}

In conclusion, organic polymer-fullerene BHJ solar cells with Lu$_3$N@C$_{80}$-PCBEH were fabricated and compared to devices with PC$_{61}$BM. The P3HT:Lu$_3$N@C$_{80}$-PCBEH devices showed a 50~\% higher $V_\text{OC}$ compared to PC$_{61}$BM based devices but a 25~\% lower $J_\text{SC}$. We investigated the reasons for the reduced current in Lu$_3$N@C$_{80}$-PCBEH based devices adressing the charge carrier photogeneration yield as well as transport properties.

As shown by XRD, AFM and TEM measurements, the intermixing in P3HT:Lu$_3$N@C$_{80}$-PCBEH is favorable for singlet exciton dissociation. However the weaker PL quenching as well the residual exciton PIA peak at 1.05~eV rather indicate less efficient charge carrier generation in P3HT:Lu$_3$N@C$_{80}$-PCBEH. Moreover, molecular triplet excitons were found to be formed in the polymer phase. We suggest the scenario of intramolecular electron transfer in Lu$_3$N@C$_{80}$-PCBEH competing with photoinduced electron transfer between polymer and fullerene. Once the CT state is formed it may decay to the lower lying triplet state via electron back transfer. Further, electrical studies allowed us to exclude that the losses of photogenerated charge carriers occurred during the transport and extraction.

Finally, we draw a physical picture on how pushing up the LUMO level of the acceptor may lead to a reduction of the overall performance of the solar cell and underline the importance of taking the relative energetics of charge transfer and triplet states into account when designing new high efficient photovoltaic BHJs.

\section{Methods}

BHJs layers for the samples and solar cells were fabricated by spin coating optical thin films on poly(3,4--ethylenedioxythiophene) poly(styrenesulfonate) covered indium tin oxide/glass substrates. In case of P3HT:Lu$_3$N@C$_{80}$-PCBEH additional slow drying was applied. Aluminum anodes were thermally evaporated. The current--voltage measurements were recorded with a Keithley 237 SMU. For the illumination of the cells we used an Oriel 1160 AM1.5G solar simulator.

For PL and PIA measurements sapphire substrates were used. P3HT was purchased from Rieke Metals and PC$_{61}$BM from Solenne. Both were used without further purification. All preparation steps were performed in a nitrogen glovebox, metal electrodes were thermally evaporated in the glove-box integrated vacuum evaporation chamber. For ESR and PLDMR measurements, the polymer  or blend solutions were poured in tubes (Wilmad), dried and sealed under vacuum to avoid oxidation of samples.

AFM and XRD measurements were performed on operating organic solar cells. AFM images were taken with a Veeco Dimension Icon and XRD curves with a GE XRD 3003 TT under ambient conditions at room temperature.

For TEM measurements, P3HT:Lu$_3$N@C$_{80}$-PCBEH layers were floated from the water-soluble poly(3,4--ethylenedioxythiophene) poly(styrenesulfonate) onto a surface of purified deionized water and picked up with 400 mesh copper TEM grids. Bright field TEM image acquisition and electron diffraction were performed on a Technai G2 20 TEM (FEI Co.), which was operated at 200kV.

For PIA spectroscopy the samples were mounted on a helium cold finger cryostat (20-293~K). During the measurement they were kept under dynamic vacuum to avoid photo-oxidation. The excitation source was a mechanically chopped cw laser at a wavelength 532~nm with a power of 68~mW. Additionally, cw-illumination was provided by a halogen lamp. Both light sources were focused onto the same spot of the sample. The transmitted light was collected by large diameter concave mirrors and focused into the entrance slit of a Cornerstone monochromator. Depending on the wavelength, the detection was provided by a silicon diode (550-1030~nm), or by a liquid nitrogen cooled InSb-detector (1030-5550~nm). Therefore, a broad energy range 0.23-2.25~eV (with the KBr cryostat windows) was accessible. The signals were recorded with a standard phase sensitive technique synchronized with the chopping frequency of the laser by using a Signal Recovery 7265 DSP lock-In amplifier. Photoinduced changes of the transmission, -$\Delta T/T$, were monitored as function of probe light wavelength. Photoluminescence measurements were done with the same set--up as used in PIA.

ESR (modified Bruker 200D) was applied to verify the presence of spin carrying polarons. The sample was placed in a resonant cavity, and cooled with a continuous flow helium cryostat, allowing a temperature range from 10~K to room temperature. The microwave absorption was measured by using lock-in, with modulation of the external magnetic field as reference. For the excitation, a 532~nm DPSS laser guided to the microwave cavity was used. The g-factor of ESR signals was calibrated for every measurement with a Bruker 035M NMR-Gaussmeter and an EIP 28b frequency counter.

For the PLDMR curves, the same setup as in ESR was used, with a sweep generator as microwave source. The microwaves ($\nu$~=~9.432~GHz), amplified by a 27dB solid state amplifier, arrive in the cavity with a power of around 60~mW. Instead of the microwave absorption, the variation of the PL intensity due to resonant microwave irradiation was recorded by using lock-In, referenced by TTL--modulating the microwave in the kHz range. All PLDMR measurements shown were recorded at T~=~10K.

Photo-CELIV measurements were performed in a closed cycle cryostat in helium atmosphere for temperatures ranging from 150 K to 300 K in steps of 25 K. There, a triangular voltage pulse in reverse direction (Agilent 81150A) is applied to the solar cell extracting free charge carriers from the bulk, whereas the charge carrier mobility and the charge carrier concentration were obtained simultaneously~\cite{Juska:2000ij}. The current transients were acquired by a digital oscilloscope (Agilent Infiniium DSO90254A) after amplification by a current--voltage amplifier (FEMTO DHPCA-100). The second harmonic of a Nd:YAG laser (532 nm, $<$ 80 ps pulse duration) was used for the laser excitation technique. The recombination dynamics was addressed by varying the delay time between the laser excitation and charge extraction. The built-in field of the solar cell was compensated by a constant offset bias during the delay time to ensure no charge loss due to extraction at the contacts.

The current transients were acquired by a digital oscilloscope (Agilent Infiniium DSO90254A) after amplification by a current--voltage amplifier (FEMTO DHPCA-100). The second harmonic of a Nd:YAG laser (532 nm, $<$ 80 ps pulse duration) was used for the laser excitation technique.


\begin{acknowledgments}
V.D.'s work at the ZAE Bayern is financed by the Bavarian Ministry of Economic Affairs, Infrastructure, Transport and Technology. A.B. thanks the German Federal Environmental Foundation (Deutsche Bundesstiftung Umwelt, DBU) for funding. C.D. acknowledges the support of the Bavarian Academy of Sciences and Humanities. A.S.'s and V.D.'s work was financially supported by the German Research Council (DFG) under contracts DY18/6-2 and INST 93/557-1.
\end{acknowledgments}

\providecommand{\WileyBibTextsc}{}
\let\textsc\WileyBibTextsc
\providecommand{\othercit}{}
\providecommand{\jr}[1]{#1}
\providecommand{\etal}{~et~al.}


\end{document}